\documentclass[preprint,12pt]{elsarticle}
\usepackage[inkscapeformat=png]{svg}
\usepackage{setspace}
\usepackage{longtable}
\usepackage{hyperref}
\usepackage{lipsum}
\usepackage{xcolor,colortbl}
\usepackage{soul}
\usepackage{lineno}

\usepackage{amssymb}

\journal{Computers \& Security}

\begin{document}

\begin{frontmatter}



\title{Assessing the Trustworthiness of Electronic Identity Management Systems: Framework and Insights from Inception to Deployment}


\author[inst2]{Mirko Bottarelli}

\affiliation[inst1]{organization={WMG, University of Warwick},
             addressline={6 Lord Bhattacharyya Way}, 
             city={Coventry},
             postcode={CV4 7AL}, 
             state={Warwickshire},
             country={UK}}

\author[inst1]{Gregory Epiphaniou}
\author[inst2]{Shah Mahmood}
\author[inst2]{Mark Hooper}
\author[inst1,inst2]{Carsten Maple}
\affiliation[inst2]{organization={The Alan Turing Institute},
            addressline={British Library, 96 Euston Rd.}, 
             city={London},
             postcode={NW1 2DB}, 
             country={UK}}

\begin{abstract}
The growing dependence on Electronic Identity Management Systems (EIDS) and recent advancements, such as non-human ID management, require a thorough evaluation of their trustworthiness.
Assessing EIDS's trustworthiness ensures security, privacy, and reliability in managing sensitive user information. It safeguards against fraud, unauthorised access, and data breaches, fostering user confidence. Existing frameworks primarily focus on specific dimensions such as security and privacy, often neglecting critical dimensions such as ethics, resilience, robustness, and reliability. 
This paper introduces an integrated Digital Identity Systems Trustworthiness Assessment Framework (DISTAF) encapsulating these six pillars. It is supported by over 65 mechanisms and over 400 metrics derived from international standards and technical guidelines. By addressing the lifecycle of DIMS from design to deployment, our DISTAF evaluates trustworthiness at granular levels while remaining accessible to diverse stakeholders. 
We demonstrate the application of DISTAF through a real-world implementation using a Modular Open Source Identity Platform (MOSIP) instance, refining its metrics to simplify trustworthiness assessment. Our approach introduces clustering mechanisms for metrics, hierarchical scoring, and mandatory criteria to ensure robust and consistent evaluations across an EIDS in both the design and operation stages. Furthermore, DISTAF is adaptable to emerging technologies like Self-Sovereign Identity (SSI), integrating privacy-enhancing techniques and ethical considerations to meet modern challenges. The assessment tool developed alongside DISTAF provides a user-centric methodology and a simplified yet effective self-assessment process, enabling system designers and assessors to identify system gaps, improve configurations, and enhance public trust.
This work advances Digital Trust by Design, laying the foundation for resilient, reliable, and ethically aligned EIDS. It enables organisations to align with regulatory requirements and operational best practices, ensuring secure and trustworthy identity management systems adaptable to future innovations.

\end{abstract}




\end{frontmatter}


\section{Introduction}
\label{sec:sample1}


Secure electronic identification is an essential component of modern cyberinfrastructures with many applications, from mobile phones and email to online shopping and e-banking. The recent pandemic increased this significance, as governments and the private sector reduced physical interactions to an absolute minimum. Electronic Identification can guarantee the unambiguous identification of a person and ensure that the correct service is provided to the rightful recipient \cite{ilie2011survey}. Nonetheless, any disruptive change to this process inevitably raises risks related to the sensitive nature of the data exchanged. It is no longer sufficient to require the security and privacy of processes; instead, one must consider whether electronic identification systems are trustworthy throughout their lifecycle from design to operation \cite{aceron2021evaluative}. Thus, understanding the different trustworthiness facets and dimensions in Electronic ID management systems (EIDS) is essential to increasing the reliability and resilience of their functions. The ability to capture, measure and assess the degree of trustworthiness against specific pillars concerned with the EIDS functions is often beyond the narrow scope of establishing and assessing a set of security requirements against legal and regulatory ramifications.

Trust requirements are often overlooked by existing and emerging digital ID assurance frameworks, especially in cases where end-to-end assurance must be established with the necessary certification and accreditation processes invoked. The emphasis is placed on the links between security control and privacy protections to enforce the requirements from both areas at different stages of the ID management lifecycle \cite{zahid2023secure}. 

Data fidelity plays an important role in establishing trustworthiness in all EIDS core operations and within the organisation since elements such as data processing credibility and precision of data-driven decision-making could broaden public acceptance and adoption \cite{zafar2017trustworthy}. Also, in any Cyber-Physical System  (CPS) that handles sensitive information, it is essential to maintain high data fidelity to ensure the integrity and reliability of that system. Poor quality and untrusted data can affect financial security and productivity, leading to erroneous decisions. This can potentially damage the trustworthiness of these systems. It is crucial to consider both resilience and reliability and other components to comprehend how entities can attain trustworthy EIDS systems. This will become even more significant with the introduction of Self-Sovereign Identity (SSI), where users will have complete control over managing their ID and technology that could automate access to services and execution of ID-related processes. Currently, no framework assesses trusted operations in SSI, with existing efforts focussing mainly on security and privacy implications alone \cite{zafar2017trustworthy}. 

This paper builds on our previous work that presents the different pillars of trustworthiness for EIDS \cite{mapleFacetsTrustworthinessDigital2021a}. We proposed an evaluation framework where trustworthiness is composed of six fundamental pillars, specifically Security, Privacy, Ethics, Resiliency, Robustness, and Reliability. Of these, 65+ primary mechanisms were identified, resulting in 400+ metrics taken from various technical reports, international standards, and guidelines. We revisit DISTAF and its metrics and assess its efficacy and effectiveness using a real-life instance of an EIDS based on MOSIP. We then simplify the number of metrics and develop a novel assessment method that enables the user to decide at which level of detail to perform an EIDS trustworthiness assessment. Some categories of users will be able to linger at higher levels, quickly capturing the coarse characteristics of the System under Testing (SuT/EIDS). 

However, designers and developers have to investigate further, diving into the details of the metrics to identify ways to improve. In both cases, it is worth noting that our method and associated tool are meant to be a companion or a reference, not to determine which EIDS implementation is better than the others, but instead, to allow each of those to gauge on their level of trustworthiness in both design and operational stages. 
Our approach involves identifying and quantifying the features of the EIDS system in two distinct stages: design and operation. This allows us to determine the degree of confidence in its ability to meet trust requirements while maintaining security, resilience, and usability. By addressing these key pillars, mechanisms, and features, we establish a foundation for developing trustworthiness assurance and self-assessment processes. This will enable the estimation and assessment of the perceived trustworthiness of EIDS systems.

The remainder of this paper is structured as follows: Section \ref{sec:relatedworks} provides related works in the field, including existing frameworks and their limitations in addressing all critical facets of trustworthiness for digital identity. Section \ref{sec:tf} presents a detailed discussion of the six pillars of trustworthiness in our DISTAF followed by the methodology for assessing trustworthiness at different stages of an EIDS (design and operation) in Section \ref{sec:am}. In Section \ref{sec:tt}, we outline the development and application of a tool for DISTAF whilst we analyse the results in Section \ref{sec:rd}. Finally, Section \ref{sec:cc} summarises our findings and future directions.


\section{Related Works}
\label{sec:relatedworks}

In this section, we situate our approach within the broader landscape of research and practice on digital identity management, highlighting the growing importance of trustworthiness and the limitations of existing frameworks in addressing all its critical facets---security, privacy, ethics, resilience, robustness, and reliability. By examining both well-established standards and emerging technologies (e.g., self-sovereign identity and privacy-enhancing mechanisms), we underscore the need for a \emph{holistic trustworthiness framework} that extends beyond mere security or privacy controls.

\subsection{Measuring Security and Privacy: Historical Context and Standards}

Historically, digital identity management practices have been shaped by standards focusing on the confidentiality, integrity, and availability (CIA) of information, as well as by data protection regulations aiming to safeguard personal data. Among the most prominent are:

\begin{itemize}
  \item \textbf{ISO/IEC 27001} (\cite{iso27001}) and \textbf{ISO/IEC 27002}, which provide comprehensive guidelines for establishing, implementing, maintaining, and continually improving an information security management system.
  \item \textbf{NIST Cybersecurity Framework} (\cite{nistCSF2018}), commonly used by organizations in the United States. A companion set of publications such as \textbf{NIST SP 800-53} (\cite{nist80053}) and \textbf{NIST SP 800-63} (\cite{nist80063}) offers detailed controls and guidelines specifically for identity proofing, authentication, and access management.
  \item \textbf{ISO/IEC 29100} (\cite{iso29100}), which defines a set of privacy principles and a high-level framework for protecting personal information.
  \item \textbf{GDPR} (\cite{gdpr2016}), the General Data Protection Regulation in the European Union, which sets stringent requirements for how personal data is collected, stored, and shared.
  \item \textbf{eIDAS (EU Regulation 910/2014)} (\cite{eidas2014}), aiming to enable seamless electronic interactions between citizens, businesses, and public authorities across EU member states.
\end{itemize}

Each of these frameworks and regulations has contributed significantly to the evolution of secure and privacy-aware identity systems. However, they tend to address security and privacy as largely distinct domains---often neglecting other essential dimensions of trustworthiness such as \emph{resilience}, \emph{ethics}, \emph{reliability}, and \emph{robustness}. For instance, while ISO/IEC 27001 prescribes critical security controls for risk assessment, it does not offer explicit guidance on ethical considerations or on building resilience into identity systems at design-time. Similarly, GDPR focuses on data protection principles but provides only limited direction on preserving system reliability or ensuring robust identity workflows under adverse conditions.

\subsection{The Need for a Comprehensive Trustworthiness Framework}

As the digital ecosystem becomes increasingly complex---relying on interoperable services and large-scale online transactions---the concept of \emph{multi-faceted trust} has emerged as a critical requirement. In the absence of a unified, high-level framework that spans \emph{security, privacy, ethics, resilience, robustness, and reliability}, implementers are left to integrate siloed guidelines themselves, an approach both fragmented and often insufficient for modern cyber-physical and distributed environments (\cite{Choo2011, Edu2023}).

The limitations become even more evident in large-scale identity solutions (e.g., national electronic identity systems, enterprise single sign-on platforms, or cross-border identity federations), where trust must be ensured continuously across diverse parties with different risk appetites, legal jurisdictions, and technological capabilities. Traditional risk management frameworks address identified threats but may struggle to incorporate ethical principles (such as fairness, transparency, and accountability) or robust system behaviors (such as continuity of service under duress) in a holistic manner.

\subsection{Trustworthiness by Design: An Integrated Approach}

In response to these shortfalls, the concept of \emph{Trustworthiness by Design (TbD)} is required, echoing earlier paradigms like \emph{Security by Design} and \emph{Privacy by Design}. TbD posits that trust considerations should be ``baked in'' to system architecture and development from the earliest stages (\cite{Cavoukian2009}). Compared to Security by Design, which largely addresses the CIA triad within system architectures (\cite{iso27001, nistCSF2018}), and Privacy by Design, which focuses on data minimization, user consent, and regulatory compliance (\cite{gdpr2016}), TbD broadens the scope.

In particular, TbD insists on integrating additional pillars such as \emph{ethics} (transparency, fairness, explainability), \emph{resilience} (the ability to adapt, detect, and recover under disruptions), \emph{robustness} (maintainability and variability control), and \emph{reliability} (repeatable and consistent behaviors). By framing trust as a multi-dimensional construct, TbD frameworks aim to systematically address both functional and non-functional requirements, including user-centric design, legal compliance, and robust service continuity under dynamic conditions (\cite{Choo2011, Cameron2005}). 

\subsection{Building on Existing Models: Integrating SSI and Blockchain for Enhanced Trust}

One of the most significant recent developments in identity management is the \emph{Self-Sovereign Identity (SSI)} model, which gives individuals full ownership and control of their credentials and personal data. Instead of relying on centralized authorities, SSI solutions often leverage \emph{blockchain} or other distributed ledgers to store cryptographic proofs related to identity (\cite{Yang2019, Allen2016}). High-profile initiatives like ID2020, Alastria, and Microsoft’s ION actively explore decentralized identity architectures that rely on transparency and user control.

Despite blockchain’s promise of immutability and decentralization of the blockchain, it does not automatically guarantee \emph{privacy} or \emph{ethical data handling}, particularly when dealing with sensitive personal information on immutable ledgers (\cite{Zyskind2015}). Integrations of \emph{Zero-Knowledge Proofs (ZKPs)}, \emph{anonymous credentials}, and \emph{selective disclosure} (\cite{Brickell2004}) can address these challenges, but large-scale adoption demands additional safeguards for regulatory compliance and interoperability. Indeed, recent studies (\cite{Krul2024}) emphasize that while SSI decentralizes trust, effective governance, robust privacy protections, and resilience against node failures remain critical.

\subsection{Privacy-Preserving Mechanisms and User-Centric Trust}

Privacy-enhancing technologies (PETs) such as \emph{homomorphic encryption}, \emph{secure multi-party computation}, and anonymous credential systems are widely recognized as enablers for privacy-first identity solutions (\cite{garcia2024}). These techniques allow verification or computation over encrypted data without revealing the data itself, thus balancing security, functionality, and user privacy.

Research also underscores that transparency and control significantly boost user trust and acceptance (\cite{Choo2011, Cavoukian2009}). From a TbD perspective, privacy-preserving mechanisms are not mere add-ons but \emph{integral} parts of organizational policies, technical architectures, and risk management strategies, ensuring holistic trustworthiness for Electronic identity management systems (EIMS).

Summarizing the above, while various security, privacy, and even decentralized identity frameworks exist, they often operate in silos or offer piecemeal solutions. This fragmentation underscores the importance of \emph{Trustworthiness by Design}, where an integrated set of pillars---security, privacy, ethics, resilience, robustness, and reliability---coalesce to form a more \emph{holistic} approach to designing and deploying digital identity management systems. In the subsequent sections, we demonstrate how our proposed trustworthiness framework addresses these pillars and introduce a systematic assessment tool to operationalize them.

\section{Trustworthiness Facets, Mechanisms and Features}
\label{sec:tf}

\begin{figure}[ht!]
\centering
\includegraphics[width=150mm]{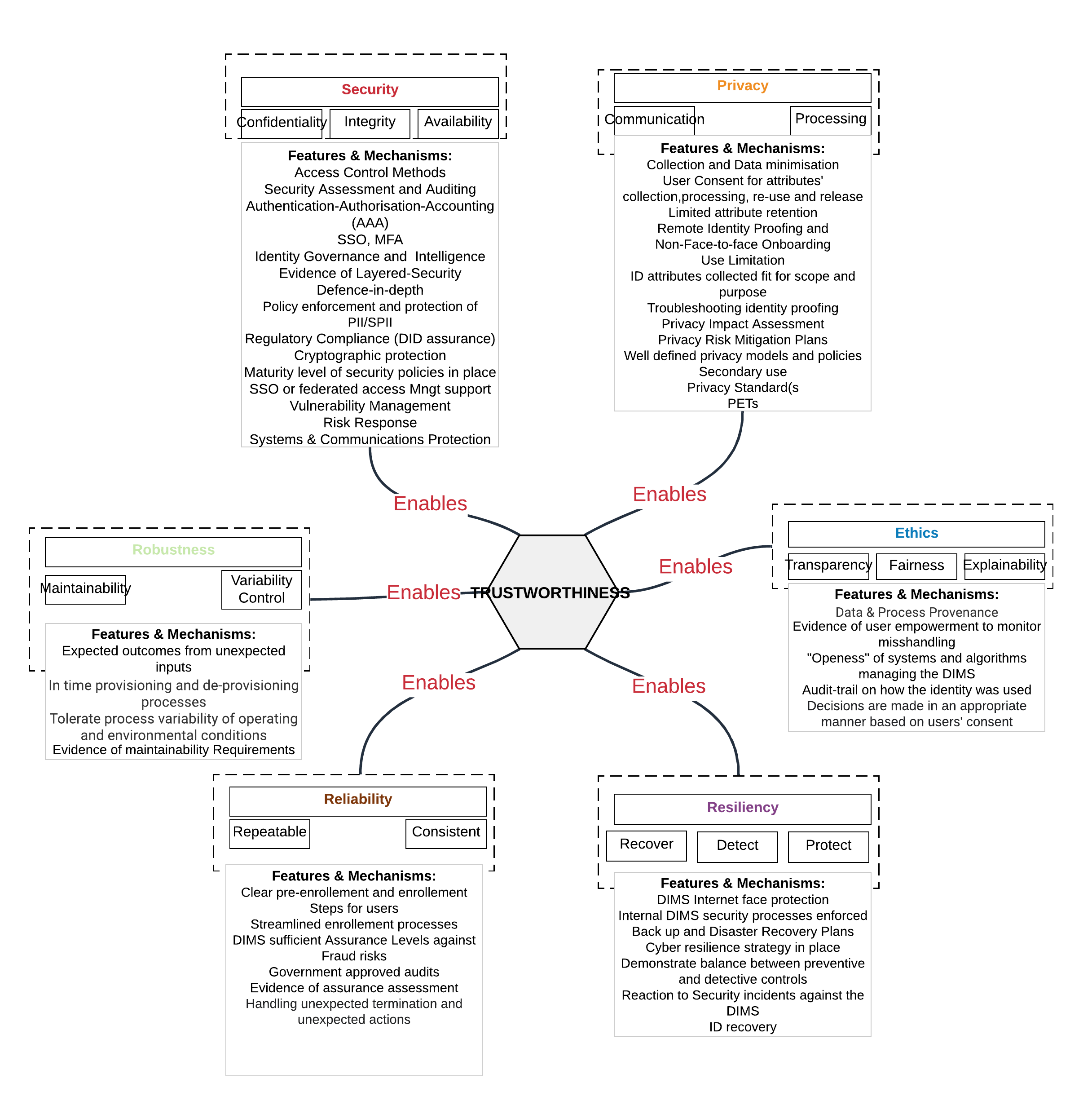}
\caption{Trustworthiness Facets, Mechanisms and Features}
\label{fig:TF}
\end{figure}

The Trustworthiness Framework \cite{mapleFacetsTrustworthinessDigital2021a} arose from the need to assess the trustworthiness assurance levels (TAL) of identity-related functions, for the certification or accreditation of EIDS systems. In fact, EIDS systems are expected to function with an acceptable risk despite any external factors, requiring a complex assessment that traditionally focusses on privacy and security requirements at different stages within the ID management lifecycle. Security involves the safeguarding of data, information, and subsystems from unauthorised access or alteration, whether in storage, processing, or during transport. Currently, privacy requirements ensure that sensitive and personal information transmitted, processed, and shared is treated privately, in adherence to the legal and regulatory restrictions that govern its use.

The current framework significantly extends the assessment, encompassing new facets that include robustness, resiliency, reliability, ethics, and their complex interplay. A trustworthiness system should be reliable, operate consistently, and in a predictable way, in accordance with its performance standards. Robustness and resiliency extend this reliability to external operating conditions. Robustness refers to the ability of the system to maintain functionality amidst internal and external difficulties without significant alterations to its original operations or condition. Likewise, resiliency emphasises the ability of the system to modify its operations in response to internal and external events, aiming to continue providing services even under altered circumstances. In a completely different plan, the ethics facet guarantees transparent, accountable, and verifiable operations throughout the whole data lifecycle.

Each of those facets, or pillars, is delineated into its corresponding properties, or mechanisms. The latter have been chosen for their relevance and importance to EIDS, with a direct impact on trust, this way influencing the trustworthiness assessment of these systems holistically. Furthermore, each mechanism is decomposed into a number of metrics that are easy to assess and whose aggregated scores could be considered as insights into the trustworthiness of the corresponding property and pillar (see Figure \ref{fig:TF}). 

\section{Trustworthiness Assessment Method for EIDS}
\label{sec:am}
The main elements of our Trustworthiness Framework are the pillars (or facets),  broken down further into mechanisms and metrics (see Figure \ref{fig:pil}). By assessing these aspects, users can accurately assess EIDS and measure the extent of their trustworthiness.  The framework presents several challenges, mainly due to the high number of metrics and the effort required for assessment. The assessment process can become quite complex with hundreds of metrics to consider. For example, the Access Control mechanism of the Security Pillar includes 16 metrics covering all aspects of such system characteristics. However, the main goal of our approach is to support users in the early stages of EIDS development. At this stage, users may only want to estimate the level of trustworthiness of an EIDS quickly. This will help assessors identify possible areas for improvement and the corresponding efforts needed. To address this need, we have implemented the following:

\begin{figure}[t!]
\centering
\includegraphics[width=100mm]{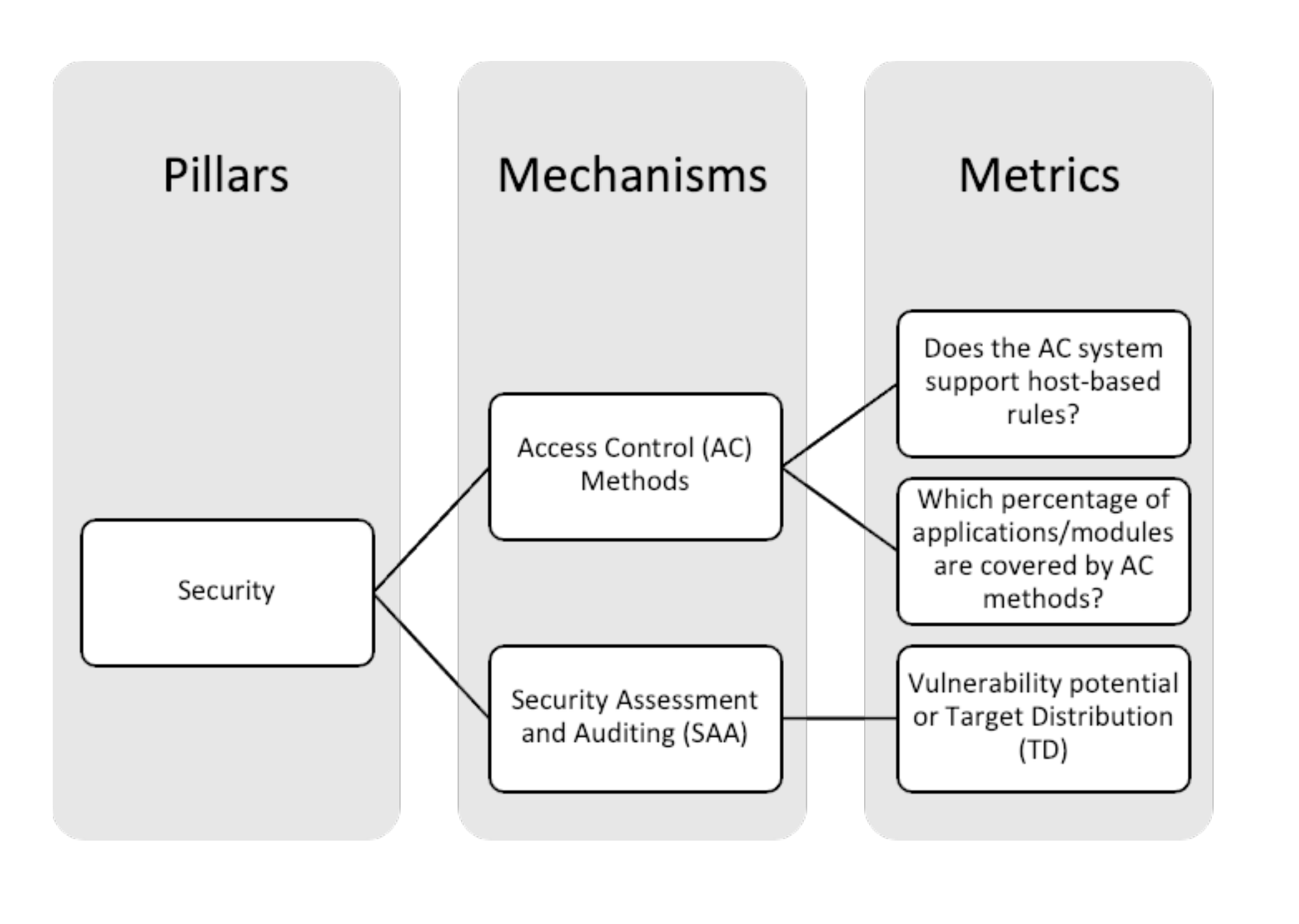}
\caption{An example of a pillar with mechanisms and metrics}
\label{fig:pil}
\end{figure}

\begin{enumerate}
\item For each mechanism, it is possible to define several common scenarios, that is, system configurations that the assessor can select to fill the underlying metrics quickly. Note that it will always be possible to change the score of individual metrics to better represent the EIDS under evaluation. The process of clustering metrics to mechanism configurations is explained in depth in the following section.
\item Since most metrics are related to standards and regulatory documents (See Table \ref{tab:standards}), the assessor can confirm their compliance, automatically providing a score to fulfilled properties.
\end{enumerate}

A second challenge arises from the amount of technicality related to some metrics, which is only suitable for some audiences, especially when the assessment must be tailored to exclude mechanisms not included in the current implementation of EIDS. To do so, each mechanism whose metrics will not be considered for the final score of the pillar can be excluded from the assessment.

To implement the three solutions, we need to identify a scoring strategy suitable at both levels, thus capable of aggregating metrics’ scores to mechanisms (and therefore to pillars), excluding those unwanted. Furthermore, the same method must also work in the opposite direction. It shall be possible to assign scores to a mechanism directly, choosing a predefined configuration with a nonambiguous way to propagate such values to the underlying metrics. In this way, no matter how the assessment is conducted, it will always be composed of a set of numerical values on the metrics, which can be used for the final evaluation, for example, by developers/designers to identify possible improvements by comparing with other systems. Furthermore, metrics need to be clustered into groups or mechanisms, resulting in a hierarchical structure. Eventually, each group will be assigned a small set of questions to allow a faster, slightly less accurate assessment. To achieve this result, we identified and executed three inner tasks described in the following sections.

\subsection{Homogeneous metrics’ scores}

In a framework of considerable scope and ambition, metrics of different types are expected to be incorporated, including scales, yes / no options, percentages, and others. Since we aim to provide an immediate overview of the characteristics of the EIDS under consideration, it is necessary to assign a single value to each pillar arising from the aggregation of the underlying mechanisms and metrics. It is therefore essential that such metrics are homogeneous. In this regard, we decided to convert all metrics to percentages and booleans, in this way without limiting the expressiveness and coverage of the assessment. 

Percentage metrics are ideal for quantifying the portion of the system subjected to external auditing or the degree of compliance to an international standard. Nevertheless, depending on the metrics, a nil percentage could indicate the ideal condition or the worst, as in the case of the vulnerable part of the system to cyberattacks and the components subject to external audit, respectively. To avoid any possible confusion, all percentage metrics range from 0\% to 100\%, with 100\% being the optimal condition. This required us to sanitise some metrics, for instance, the False Rejection Rate metric, as depicted in Table \ref{tab:example-metric-sanitised}.

\begin{table}[ht!]
  \centering
  \begin{tabular}{|p{3cm}|p{9.5cm}|}
    \hline
    \textbf{Metric} & S.SAA.O10  \\ \hline
    \textbf{Description}   & False Rejection Rate (FRR) is the probability (0-100\%) that the AAA system does not detect a match between the user-provided input and the corresponding entry in the database. Or, equivalently, it estimates the percentage of valid inputs that are incorrectly rejected.    \\ \hline
    \textbf{Pillar \newline $\rightarrow$ Mechanism}   & Security \newline $\rightarrow$ Authentication Authorization Accounting (AAA)   \\ \hline
    \textbf{Phase} & Operational  \\ \hline
    \textbf{Score} & 100\%-FRR \\ \hline
  \end{tabular}
  \caption{Example of a metric whose score was sanitised in order to make it homogeneous, thus aggregable with the rest of the mechanism's properties.}
  \label{tab:example-metric-sanitised}
\end{table}

Boolean metrics, which only accept Yes or No values, were used to declare whether a particular property was considered or implemented in the considered EIDS. Under the hood, Boolean scores are converted to percentages, 100\% if true or 0\% otherwise, allowing their aggregation.

\subsection{Clustering metrics in mechanisms}

During some initial tests, it emerged that the direct aggregation of metrics in the various pillars was not easy, both because of the high number of metrics to consider and because of the cognitive distance between a system property, sometimes even technical, and much more abstract concepts represented by the pillars. It therefore seemed reasonable to create a grouping of metrics, the mechanisms, capable of bridging this gap.

The clustering process was challenging due to the quantity and diversity of the metrics involved. An iterative approach was essential, necessitating a meticulous review of the results to identify the most relevant clusters and accurately classify them. The methodology employed was based on thematic analysis \cite{braun2006using}, borrowed from psychology. During the preliminary phase, all metrics were rewritten (encoded) to use the same terminology, highlighting the intersections and interconnections among them. Subsequently, first-level groups were identified, carefully aiming to minimise the inherent accuracy loss that is unavoidable in any clustering process. The underpinning idea was to group metrics that relate to the same notion or element of the system under evaluation. Moreover, mechanisms where defined solely when it was feasible to devise a question, typically multiple-choice, whose responses could be depicted as configurations of the original metrics. Nonetheless, there could be instances where the multiple-choice question encompasses all conceivable combinations of the underlying metrics, hence complicating the evaluation. In these cases, we exclusively consider the most common circumstances, leaving the user the ability to drill down to the metric level for finer granularity.  

As an example, consider the metrics shown in Table \ref{tab:example-metrics-clustering}. The top and middle metrics are related to the configurability of Access Control (AC) systems. Furthermore, they both are assessed during the design phase, making them good candidates to become a cluster, as depicted at the bottom of the same table. Each possible answer also provides the corresponding configuration of the underlying metrics, which is needed for hierarchical score propagation.  

\begin{table}[ht!]
  \centering
  \begin{tabular}{|p{3cm}|p{9.5cm}|}
    \hline
    \textbf{Metric} & S.AC.D8  \\ \hline
    \textbf{Description}   & Support for host-based rules.   \\ \hline
    \textbf{Pillar \newline $\rightarrow$ Mechanism}   & Security \newline $\rightarrow$ Access Control (AC) Methods   \\ \hline
    \textbf{Phase} & Design  \\ \hline
    \textbf{Score} & Y/N  \\ \hline
  \end{tabular}

\vspace{0.5cm}

  \begin{tabular}{|p{3cm}|p{9.5cm}|}
    \hline
    \textbf{Metric} & S.AC.D9  \\ \hline
    \textbf{Description}   & Flexibility to handle security-policy changes   \\ \hline
    \textbf{Pillar \newline $\rightarrow$ Mechanism}   & Security \newline $\rightarrow$ Access Control (AC) Methods   \\ \hline
    \textbf{Phase} & Design  \\ \hline
    \textbf{Score} & 0 (0\%) - Not configurable \\
    & 1 (25\%) - Limited flexibility   \\
    & 2 (50\%) - Standard flexibility   \\
    & 3 (75\%) - Role-based or advanced configuration   \\
    & 4 (100\%) - Handles possible future policy changes   \\ \hline 
  \end{tabular}

\vspace{0.5cm}
    $\Downarrow$ clustering
\vspace{0.5cm}
  
  \begin{tabular}{|p{3cm}|p{9.5cm}|}
    \hline
    \textbf{Cluster} & AC Configurability   \\ \hline
    \textbf{Question}   & What is the level of AC configurability?   \\ \hline
    \textbf{Pillar \newline $\rightarrow$ Mechanism}   &     Security \newline $\rightarrow$ Access Control (AC) Methods   \\ \hline
    \textbf{Phase} & Design  \\ \hline
    \textbf{Answers} & a) limited or no configurability \newline $\longrightarrow$ (S.AC.D8=N, S.AC.D9=0\%)  \\
    & b) simple level of configurability \newline $\longrightarrow$ (S.AC.D8=N,  S.AC.D9=25\%)    \\
    & c) advanced configuration \newline $\longrightarrow$ (S.AC.D8=N, S.AC.D9=75\%)    \\
    & d) support for current and future policies \newline $\longrightarrow$ (S.AC.D8=Y, S.AC.D9=100\%)    \\ \hline
  \end{tabular}

  \caption{Example of access-control metrics suitable for clustering, and the resulting group with associated question. Note how each group answer provides the scores of the underlying metrics.}
  \label{tab:example-metrics-clustering}
\end{table}

\subsection{Hierarchical scoring}

\begin{figure}[t!]
\centering
\includegraphics[width=100mm]{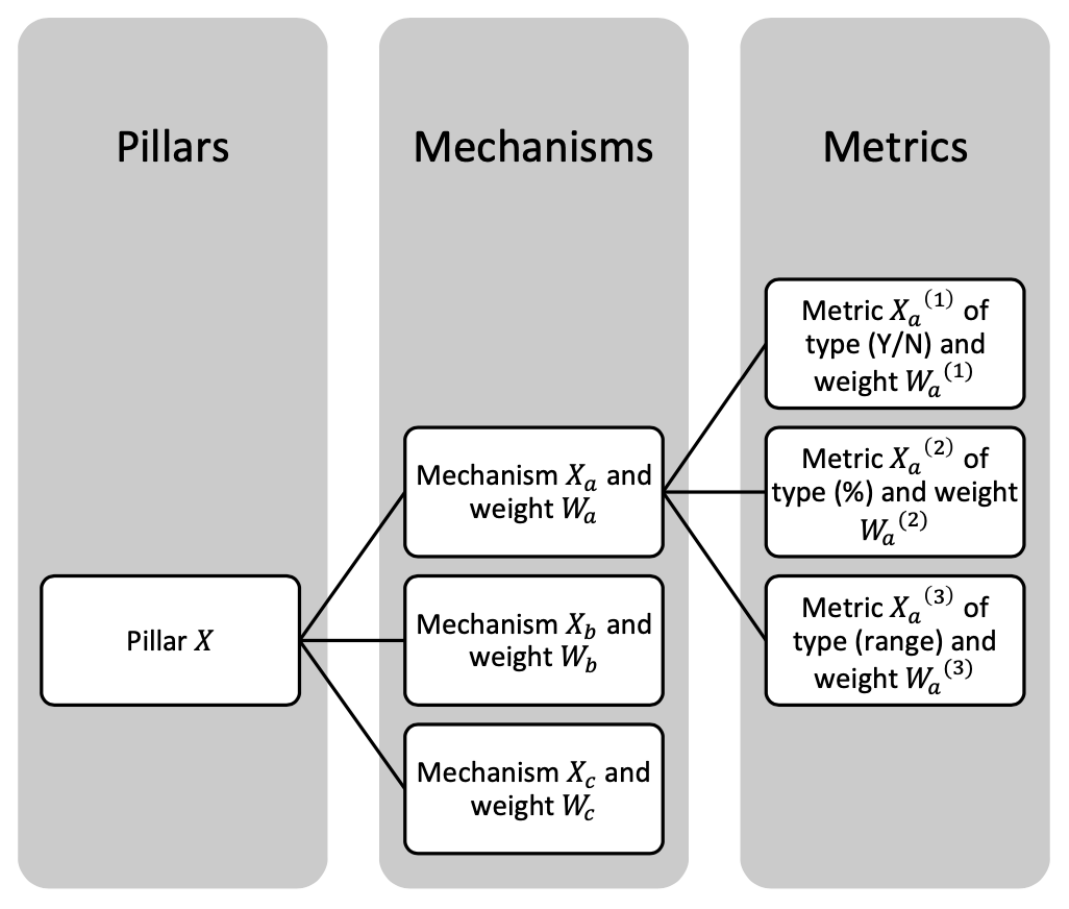}
\caption{A generalisation of a pillar-mechanism-metrics hierarchy for score propagation.}
\label{fig:hierarchical_scoring}
\end{figure}

In order to come up with a unique score for each trustworthy facet of an EIDS, it is necessary to propagate metric scores to the root of each pillar-mechanisms-metric hierarchy. Figure \ref{fig:hierarchical_scoring} illustrates a generic structure where pillar $X$ is composed of $|X|=3$ mechanisms $X_a, X_b, X_c$ with related weights $W_a, W_b, W_c$. Each mechanism is, in turn, composed of a number of metrics, for instance, mechanism $X_a$ is composed of $|X_a|=3$ metrics, defined as $X_a^{(1)}, X_a^{(2)}, X_a^{(3)}$ with corresponding weights $W_a^{(1)}, W_a^{(2)}, W_a^{(3)}$. Weights allow modulating the significance of mechanisms and metrics. By default, they have a value of $1/n$, treating all mechanisms (and metrics) as equally important. 

Scores are propagated according to the following three steps:
\begin{enumerate}
    \item \textbf{Top-down propagation from mechanisms to metrics}. In the case where the assessor decides to use cluster questions, it is necessary to propagate the score to the underlying metrics. In detail, for each pillar $X$, if the mechanism $X_y$ was assessed through the multiple choice question, then all metrics $X_y^{(i)}, i\in\{1...|X_y|\}$ are scored according to the configuration declared in the chosen answer.
    \item \textbf{Bottom-up aggregation from metrics to mechanisms}. For each pillar $X$, the score for each mechanism $X_y$ is the weighted sum of its metrics, thus:
    $$score(X_y)=\sum_{i=1}^{|X_y|} score(X_y^{(i)})\cdot W_y^{(i)}$$
    \item \textbf{Bottom-up aggregation from mechanisms to pillars}. As the final step, mechanisms' scores are aggregated to the corresponding pillar, as follows:
    $$score(X)=\sum_{i=1}^{|X|} score(X_y)\cdot W_y$$
\end{enumerate}
Note that pillars are not aggregated as they provide entirely different perspectives on the system.

\subsection{Mandatory metrics}

The aggregation of metrics is a powerful strategy, especially if weights could be used to define which metrics have more impact than others. However, it is still possible to achieve a medium-high score for a mechanism and a pillar, despite failure to meet one or more critical criteria. For instance, one could claim that the automated password recovery system, described in the operational metric RES.IDR.O6, is essential for its corresponding mechanism, under the resilience pillar. Without this metric being met, the mechanism cannot be deemed adequately designed or implemented. 

To fulfil this necessity, the current framework provides the possibility of specifying some metrics as mandatory. Furthermore, the mandatory property is not defined as a simple Boolean value; instead, each metric defines two caps, one for the corresponding mechanism and the other for the corresponding pillar. The rationale is that if such a metric is not satisfied, the mechanism and/or the pillar cannot receive a score greater than the related caps, giving much flexibility to the framework itself. Continuing the example provided above, the RES.IDR.O6 metric has been defined with caps 40\%-80\%, indicating that if not met, the maximum score at the mechanism is restricted to 40\%, while the cumulative score at the pillar level is restricted to 80\%. 

In case of multiple mandatory metrics, the mechanism and the pillar caps are defined as the minimum values of the corresponding caps defined on the metrics. Symmetrically, declaring a cap of 100\% indicates that the metric is not mandatory. As a result, the capped (final) scores for mechanisms and pillars are defined as:

$$capscore(X_y)=\min(score(X_y), \min_{1\leq i \leq |X_y|}cap(X_y^{(i)}))$$
$$capscore(X)=\min(score(X), \min_{1\leq i \leq |X_y|, 1\le y \le |X|}cap(X_y^{(i)}))$$

\noindent where $cap(\cdot)$ returns the mechanism-pillar caps of a metric if not satisfied, otherwise 100\%.

The introduction of mandatory metrics required the implementation of another functionality, that is, the possibility of excluding mechanisms from the assessments. In fact, there are cases where the system under analysis cannot be evaluated according to a specific trustworthy aspect because the latter will never be implemented. Leaving such a mechanism enabled in the assessment may inevitably cap the overall score, especially in the case where it contains mandatory metrics.


\section{Trustworthiness Assessment Tool}
\label{sec:tt}

Even if the proposed framework could be used in the form of a spreadsheet file, we decided to design and implement a web-based application to guide the users, simplifying and speeding up the assessment procedure. Users fall into three main roles:

\begin{description}
    \item[Admins] are responsible for managing the credentials and roles of other users. They are capable of adding new users to certain roles, disabling access, and regenerating temporary passwords.
    \item[Assessors] manage assessments, both by accessing existing ones and by creating a new one. The latter can originate from a blank template or as a revised version of a previous entry. In addition, the assessors actually lead the assessments by assigning scores to the metrics.
    \item[Externals] can access the assessments in a read-only fashion, for example, to export the results, in a variety of formats.
\end{description}

\begin{figure}[t!]
\centering
\includegraphics[width=9cm]{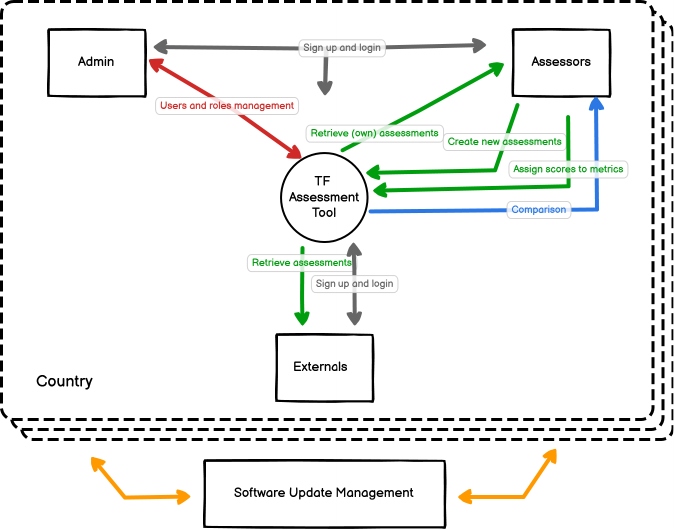}
\caption{The context diagram illustrates the actors and the corresponding interactions with the proposed tool.}
\label{fig:context-diagram}
\end{figure}

The diagram in Figure \ref{fig:context-diagram} illustrates the context of the tool, along with all actions that could be executed by each role.

\subsection{Managing assessments}

The heart of the tool is the management of assessments, each of which is composed of a unique code, a description, creation date, last modification, and status. The latter can be draft, private and public. The draft status is the one set by default when creating a new entry, and indicates that the assessment is still to be concluded. When completed, the assessments can become private or public, where the former is only visible to users with the assessor role, while the public ones are also accessible in read-only fashion by external users.

There are two ways to create a new assessment: a) from scratch and b) from an existing assessment. In the first case, the assessment is created starting from a completely empty master template, provided with the tool itself, containing the entire hierarchy of pillars, mechanisms, and metrics. In the second case, the new assessment inherits all the scores of its predecessor, allowing, for instance, the system integrator to re-evaluate the trustworthiness of its system after releasing a new release.

\begin{figure}[t!]
\centering
\includegraphics[width=12cm]{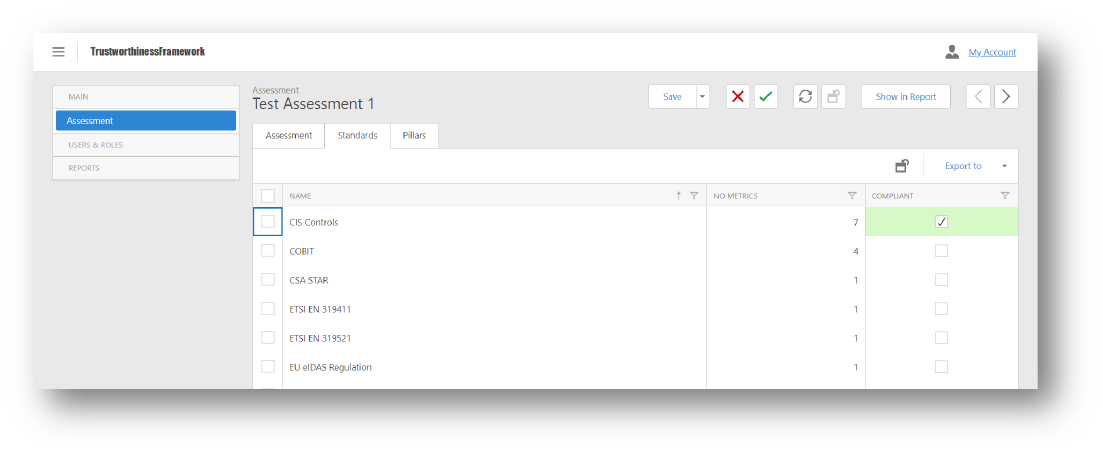}
\caption{Several standards are included in the framework, which allow for quick assessment of related metrics.}
\label{fig:tf-edit-standards}
\end{figure}

As stated above, assessments can be performed at different levels of detail. Since several metrics are derived from standards and regulatory documents, it is sufficient to indicate compliance with these to automatically assign the right score to the metrics involved. For example, Figure \ref{fig:tf-edit-standards} shows how compliance with CIS Controls automatically satisfies seven metrics.

\begin{figure}[t!]
\centering
\includegraphics[width=12cm]{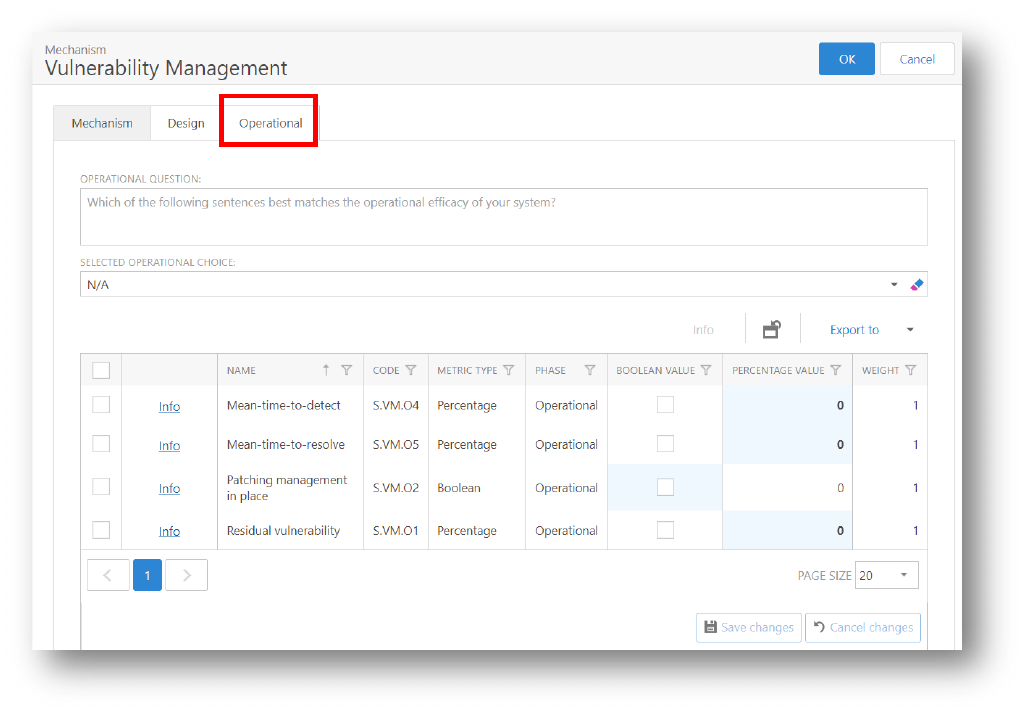}
\caption{Scores may be assigned starting from the corresponding mechanisms, either by multiple choices questions or by directly inputting metrics.}
\label{fig:tf-edit-mechanisms}
\end{figure}

For finer granularity, the assessor can choose the pillar and therefore the mechanism on which to operate, as illustrated in Figure \ref{fig:tf-edit-mechanisms}. Each mechanism can be evaluated in the design phase and in the operational phase. For both, a multiple choice question is available that simplifies and makes the assignment of scores to metrics immediate under the most common conditions. It is still possible to define a specific score for each metric, whose description and references are provided contextually.

\subsection{Assessment reports}

\begin{figure}[t!]
\centering
\includegraphics[width=12cm]{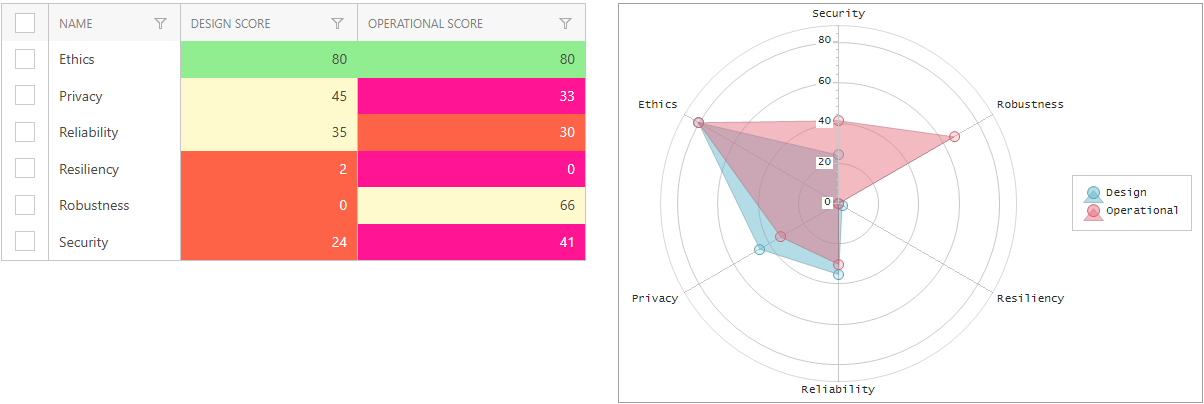}
\caption{Trustworthiness fingerprint of EIDS under testing.}
\label{fig:tf-fingerprint}
\end{figure}

Once the assessment is completed, it is essential that its vision is effective in suggesting, in a simple way, the most promising directions for improvement. For example, if the security pillar has a very low score, it must be immediate to understand which mechanisms, and therefore which metrics, have led to this result. To achieve this goal, the tool uses several techniques. Firstly, Table \ref{tab:colours} lists all the colours used as the background of all scores, at each level of the hierarchy. In addition, the tools provides useful polar diagrams at both the mechanisms and pillar levels, for both the design and the implementation phases. These charts provide a quick overview of the assessment, as a sort of trustworthiness fingerprints of the entire system and each of its components (Figure \ref{fig:tf-fingerprint}).

\definecolor{deeppink}{RGB}{255, 20, 147}
\definecolor{tomatored}{RGB}{255,99,71}
\definecolor{lemonchiffon}{RGB}{255, 250, 205}
\definecolor{lightgreen}{rgb}{0.56, 0.93, 0.56}

\begin{table}[ht!]
  \centering
  \begin{tabular}{|p{3cm}|p{9.5cm}|}
    \hline
    \cellcolor{deeppink}\color{white}Deep Pink & When applied to a metric, it means such metric is mandatory and it forces a cap at the score of the corresponding mechanism and/or pillar. When applied to a mechanism or a pillar, it means that one ore more mandatory metrics are not satisfied.  \\ \hline
    \cellcolor{tomatored}\color{white}Tomato Red & When applied to a metric, a mechanism, and a pillar, it indicates that the corresponding score is $\le 33\%$.  \\ \hline
    \cellcolor{lemonchiffon}Lemon Chiffon & When applied to a metric, a mechanism, and a pillar, it indicates that the corresponding score is between $33\%$ and $66\%$. \\ \hline
    \cellcolor{lightgreen}Light Green & When applied to a metric, a mechanism, and a pillar, it indicates that the corresponding score is $> 66\%$. \\ \hline
    Transparent & Used at the mechanism level indicates that such a mechanism has been excluded from the assessment. \\ \hline   
  \end{tabular}
  \caption{Colour table use for score backgrounds.}
  \label{tab:colours}
\end{table}

The tool also allows for comparing assessments; however, such functionality is only available to assessors and not to externals, which encourages the use of this tool to support the design and implementation of better EIDSs instead of comparing different products.

\subsection{Implementation and testing of DISTAF against a MOSIP Instance}
\label{sec:rd}
To validate and refine our DISTAF, we constructed a controlled experimental environment using a MOSIP-based Electronic Identity System (EIDS), referred to as \textbf{Turing’s ID system}. This system was intentionally kept simple, excluding advanced operational components such as Security Operations Center (SOC) integration, continuous threat monitoring, and advanced resilience mechanisms. The purpose was to create a minimal yet functional identity system that could serve as a reliable testbed for DISTAF's mechanisms and metrics.

The experiment was conducted in two distinct phases:
\begin{enumerate}
    \item \textbf{Design Phase Assessment:} This phase involved evaluating the architectural and policy-level mechanisms of the system before deployment.
    \item \textbf{Operational Phase Assessment:} This phase assessed the real-time operational performance, focusing on incident handling, service continuity, and user-centric features.
\end{enumerate}

Both phases were assessed using the DISTAF tool, enabling structured scoring and result aggregation.

\begin{figure}[t!]
    \centering
    \begin{tabular}{ccc}
        \begin{minipage}[t]{0.3\linewidth}
            \centering
            \includegraphics[width=\linewidth]{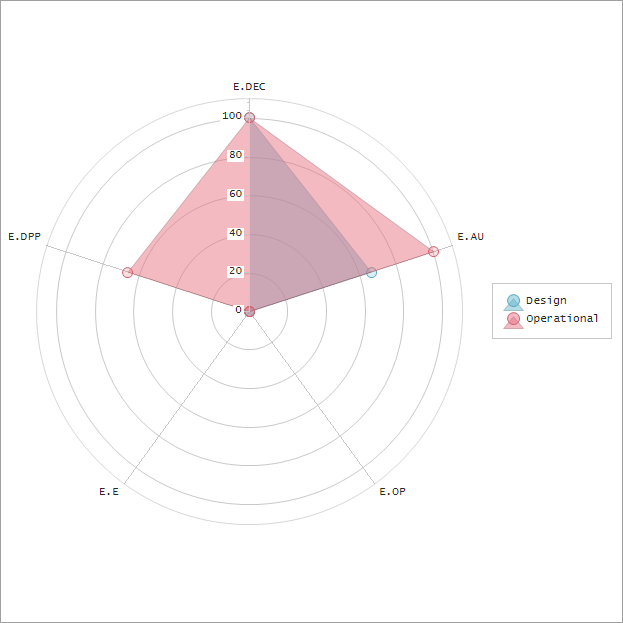}
            \caption{Turing's EIDS Performance in the Ethics Pillar}
            \label{fig:TuringIDS_EthicsPillar}
        \end{minipage} &
        \begin{minipage}[t]{0.3\linewidth}
            \centering
            \includegraphics[width=\linewidth]{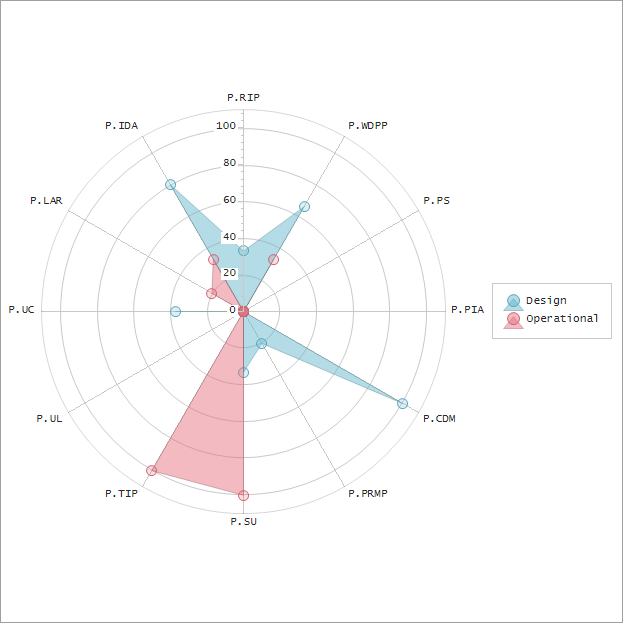}
            \caption{Turing's EIDS Performance in the Privacy Pillar}
            \label{fig:TuringIDS_PrivacyPillar}
        \end{minipage} &
        \begin{minipage}[t]{0.3\linewidth}
            \centering
            \includegraphics[width=\linewidth]{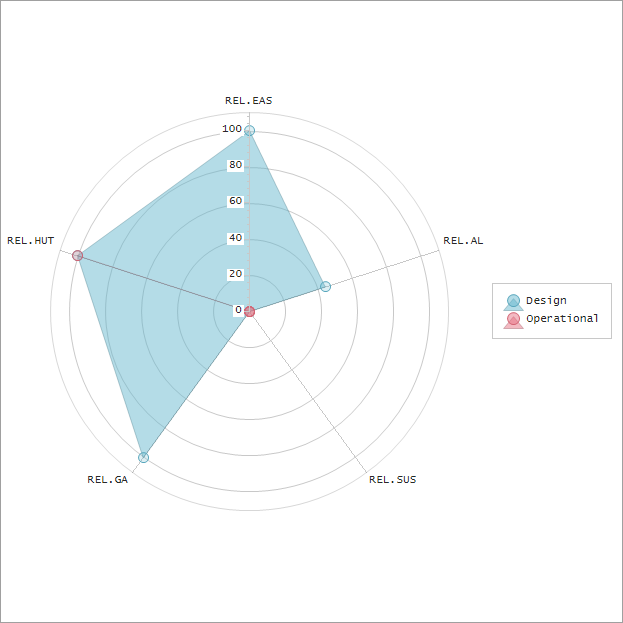}
            \caption{Turing's EIDS Performance in the Reliability Pillar}
            \label{fig:TuringIDS_ReliabilityPillar}
        \end{minipage} \\
        \begin{minipage}[t]{0.3\linewidth}
            \centering
            \includegraphics[width=\linewidth]{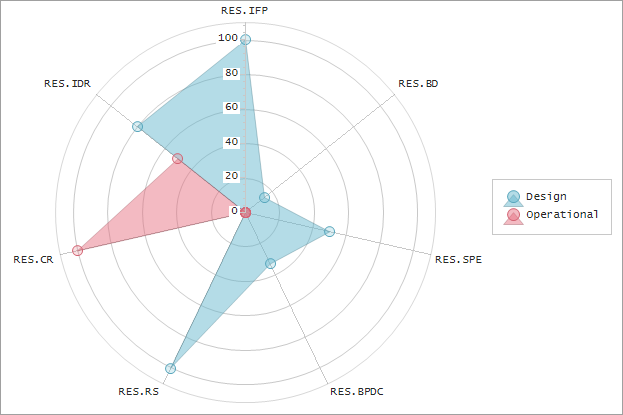}
            \caption{Turing's EIDS Performance in the Resiliency Pillar}
            \label{fig:TuringIDS_ResiliencyPillar}
        \end{minipage} &
        \begin{minipage}[t]{0.3\linewidth}
            \centering
            \includegraphics[width=\linewidth]{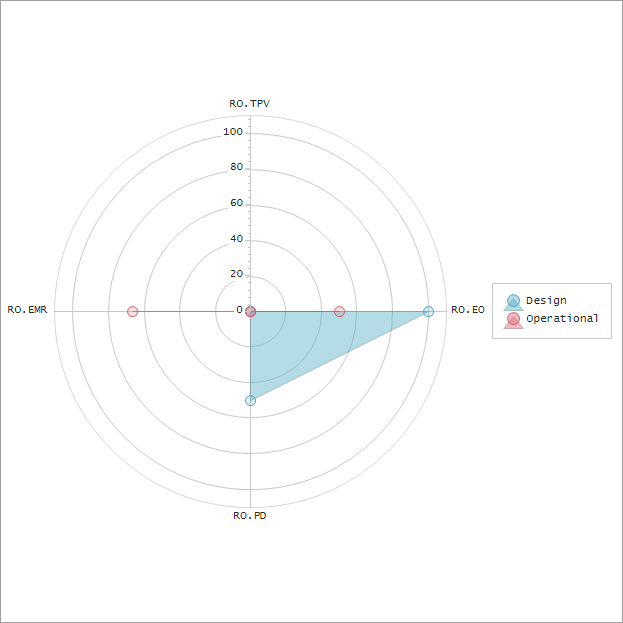}
            \caption{Turing's EIDS Performance in the Robustness Pillar}
            \label{fig:TuringIDS_RobustnessPillar}
        \end{minipage} &
        \begin{minipage}[t]{0.3\linewidth}
            \centering
            \includegraphics[width=\linewidth]{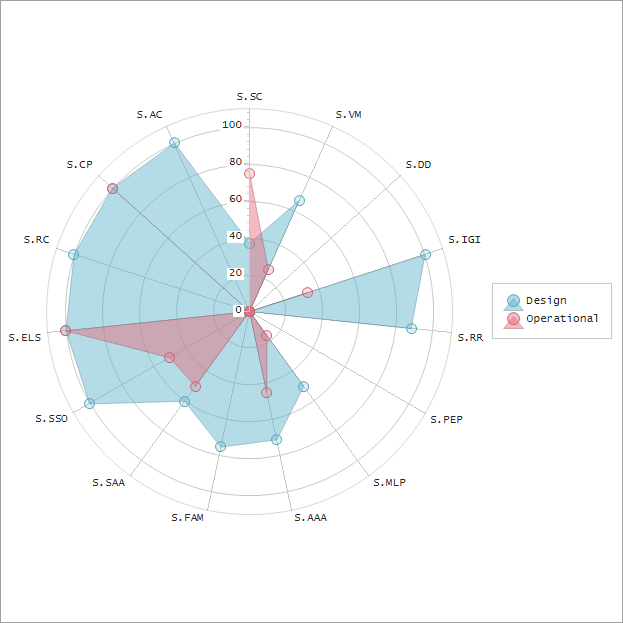}
            \caption{Turing's EIDS Performance in the Security Pillar}
            \label{fig:TuringIDS_SecurityPillar}
        \end{minipage}
    \end{tabular}
\end{figure}

\begin{figure}[t!]
\centering
\includegraphics[width=8cm]{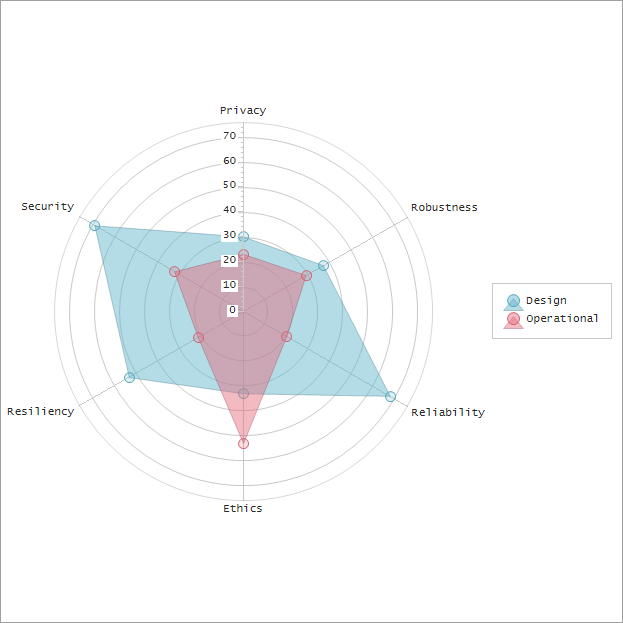}
\caption{Turing's EIDS Performance across all 6 pillars of Trustworthiness}
\label{fig:TuringIDS_Trustworthiness}
\end{figure}

Since Turing’s ID system was intentionally designed as a basic testbed, many advanced cybersecurity mechanisms present in real-world EIDS deployments—such as SOC integration, adaptive threat detection, and automated incident response-were excluded. Consequently, the overall scores are on the lower end, reflecting the simplicity of the system rather than any fundamental deficiencies in DISTAF. In the Figures \ref{fig:TuringIDS_EthicsPillar},\ref{fig:TuringIDS_PrivacyPillar},\ref{fig:TuringIDS_ReliabilityPillar},\ref{fig:TuringIDS_ResiliencyPillar},\ref{fig:TuringIDS_RobustnessPillar},\ref{fig:TuringIDS_SecurityPillar} below please find the performance of Turing's ID system across the different pillars. Also in Figure \ref{fig:TuringIDS_Trustworthiness} you can see the overall performance of Turing's IDS across the 6 pillars and both phases. 

\section{Conclusion}
\label{sec:cc}
We present the Digital Identity Systems Trustworthiness Assessment Framework (DSTAF) and its companion tool here as the first attempt to provide a comprehensive methodology for self-assessing the trustworthiness of EIDS throughout their entire lifecycle, from conception to operation. We uniquely integrate 65 mechanisms and define over 400 metrics by using a diverse set of standardisation documents and best practice guidelines and ensuring rigorous multi-dimensional assessments across six pillars: Security, Resilience, Privacy, Ethics, Robustness and reliability. Our approach distinguishes and tailors the assessment metrics for both the design and operational stages, allowing auditors and practitioners to use DISTAF to reflect their current EIDS deployment status and performance across trustworthiness dimensions. It supplements existing self-assessment efforts in preparation for conformance testing across these dimensions and offers a structured approach for identifying gaps and areas of improvement, with a dedicated web-based tool supporting this process. DISTAF's effectiveness was validated by applying it to a real-world instance of MOSIP in a controlled environment. As part of our future work, we aim to expand DISTAF's capabilities to support in-country testing and deployment, refining the tool and metrics to align with constantly evolving legal and regulatory compliance requirements and ramifications.


\appendix

\section{List of Documentation}
\label{sec:sample:appendix}

\begin{longtable}{| p{.20\textwidth} | p{.80\textwidth} |} 
\hline
    \textbf{Sample Metric Codes} & \textbf{Reference Standards / Guidelines}  \\ \hline
   S.AC.D1 &	ISO/IEC 27001:2022 (with Amendment 1:2024), NIST SP 800-53 Rev. 5, ISO 22301:2019, ISO 31000:2018, IEC 62645:2014 \\ \hline
S.AC.D2	& ISO/IEC 27001:2022 (with Amendment 1:2024), NIST SP 800-53 Rev. 5, ISO 22301:2019, ISO 31000:2018, ISO/IEC 27002:2022 \\ \hline
S.AC.D5 &	\href{https://csrc.nist.gov/glossary/term/policy_enforcement_point}{Policy Enforcement} \\ \hline
S.AC.D7	& ISO/IEC 27001:2022 (with Amendment 1:2024), NIST SP 800-53 Rev. 5, ISO/IEC 27002:2022, ISO/IEC 27005:2018, COBIT 2019\\ \hline
S.AC.D10 &	AC-12 SESSION TERMINATION from NIST 800-171\\ \hline
S.AC.O6	& NISTIR 7874\\ \hline
S.AC.O7	& NISTIR 7875\\ \hline
S.AC.O8	& NISTIR 7876\\ \hline
S.SAA.D1	& ISO/IEC 27001:2022 (with Amendment 1:2024), ISO/IEC 27002:2022, NIST SP 800-53 Rev. 5, ISO/IEC 27005:2018, COBIT 2019\\ \hline
S.SAA.D2 &	MAGERIT, MITRE CVSS\\ \hline
S.SAA.D5 &	NIST 800-137\\ \hline
S.SAA.D6 &	ISO/IEC 27001:2022 (with Amendment 1:2024), ISO/IEC 27002:2022, ISO/IEC 27005:2018, NIST SP 800-53 Rev. 5, COBIT 2019\\ \hline
S.SAA.D7 &	SO/IEC 27001:2022 (with Amendment 1:2024), ISO/IEC 27002:2022, ISO/IEC 27005:2018, NIST SP 800-53 Rev. 5, COBIT 2019\\ \hline
S.AAA.D6 &	NIST 800-12\\ \hline
S.AAA.D7 &	NIST 800-192\\ \hline
S.PEP.D1 &	NIST 7874\\ \hline
S.RC.D1 &	ISO 19011 (Guidelines for auditing management systems)\\ \hline
S.RC.D3 &	ISO/IEC 27001 (Information Security Management Systems)\\ \hline
S.RC.O8 &	ISO270001\\ \hline
S.RC.O9 &	ISO24760\\ \hline
S.RC.O10&	ISO29115\\ \hline
S.RC.O11&	ISO29146\\ \hline
S.RC.O12&	GDPR\\ \hline
S.RC.O13&	ETSI EN 319411\\ \hline
S.RC.O14&	ETSI EN 319521\\ \hline
S.CP.D3&	NIST 800-57\\ \hline
S.MLP.D7&	NIST 800-128\\ \hline
S.RR.D1&	MAGERIT methodology\\ \hline
S.RR.D4&	ISO27005 InfoSec Risk Management, SP800 37R2 \\ \hline
S.RR.O1&	MAGERIT \\ \hline
S.RR.O15&	ISO 27001:2013\\ \hline 
S.SC.D3&	NIST SP800-53(r4)\\ \hline
S.SC.D4&	SP 800-41 Rev. 1, Guidelines on Firewalls and Firewall Policy\\ \hline
S.SC.D5&	NIST Publishes SP 800-189, Resilient Interdomain Traffic Exchange: BGP Security and DDoS Mitigation\\ \hline
S.SC.D6&	Guide to IPsec VPNs\\ \hline
S.SC.D7&	Guide to Intrusion Detection and Prevention Systems (IDPS)\\ \hline
S.SC.O1&	NIST SP 800-53 (Security and Privacy Controls for Federal Information Systems and Organizations) - Controls such as AC-17 (Remote Access) and SC-8 (Transmission Confidentiality and Integrity)\\ \hline
S.SC.O2&	NIST SP 800-53 (Security and Privacy Controls for Federal Information Systems and Organizations) - Controls such as SC-7 (Boundary Protection) and SC-32 (Information System Partitioning)\\ \hline
S.SC.O3&	NIST SP 800-53 (Security and Privacy Controls for Federal Information Systems and Organizations) - Control AC-5 (Separation of Duties) and SC-2 (Application Partitioning)\\ \hline
S.SC.O4&	NIST SP 800-53 (Security and Privacy Controls for Federal Information Systems and Organizations) - Control SC-7 (Boundary Protection)\\ \hline
S.SC.O5&	NIST SP 800-61\\ \hline
S.SC.O6&	NIST SP 800-53 (Security and Privacy Controls for Federal Information Systems and Organizations) - Controls such as SC-7 (Boundary Protection) and AC-17 (Remote Access)\\ \hline
S.SC.O8&	OWASP (Open Web Application Security Project) Guidelines\\ \hline
S.SC.O9&	NIST SP 800-53 (Security and Privacy Controls for Federal Information Systems and Organizations) - Controls such as IA-5 (Authenticator Management) and SC-12 (Cryptographic Key Establishment and Management)\\ \hline
P.CDM.D1&	GDPR (General Data Protection Regulation)\\ \hline
P.CDM.O1&	ISO/IEC 25010 (System and software quality models)\\ \hline
P.CDM.O2&	ISO/IEC 25012 (Data Quality Model)\\ \hline
P.CDM.O3&	NIST SP 800-53 (Security and Privacy Controls for Federal Information Systems and Organizations)\\ \hline
P.UC.D3&	Data Retention and Investigatory Powers Act 2014 \\ \hline
P.UC.D4&	ISO 10003 (Quality management -- Customer satisfaction -- Guidelines for dispute resolution external to organizations)\\ \hline
P.UC.D7&	NIST SP 800-88 (Guidelines for Media Sanitization)\\ \hline
P.UC.D8&	Data Retention Regulations 2014\\ \hline
P.UC.O1&	ISO/IEC 27031 (Guidelines for information and communication technology readiness for business continuity)\\ \hline
P.UC.O4&	SO 10003 (Quality management -- Customer satisfaction -- Guidelines for dispute resolution external to organizations)\\ \hline
P.UC.O7&	\href{https://ico.org.uk/for-organisations/guide-to-data-protection/guide-to-le-processing/individual-rights/the-right-to-be-informed/}{Right to be Informed}\\ \hline
P.LAR.D1&	ISO 15489 (Information and Documentation - Records Management)\\ \hline
P.LAR.O7&	\href{https://en.wikipedia.org/wiki/Data_Protection_Directive}{Data Protection Directive}\\ \hline
P.RIP.D1&	eIDAS (Electronic Identification and Trust Services)\\ \hline
P.RIP.D2&	NIST SP 800-63A (Digital Identity Guidelines: Enrollment and Identity Proofing)\\ \hline
P.RIP.D3&	ISO/IEC 29115 (Entity Authentication Assurance Framework)\\ \hline
P.RIP.D4&	SO/IEC 29115 (Entity Authentication Assurance Framework)\\ \hline
P.RIP.O2&	NIST SP 800-137 (Information Security Continuous Monitoring (ISCM) for Federal Information Systems and Organizations)\\ \hline
P.RIP.O3&	ISO/IEC 27001:2013 \\ \hline
P.UL.D1&	SO/IEC 29100 (Privacy Framework)\\ \hline
P.UL.O1&	ISO/IEC 29100 (Privacy Framework)\\ \hline
P.IDA.D2&	NIST SP 800-63B (Digital Identity Guidelines: Authentication and Lifecycle Management) for identity lifecycle management including revocation.\\ \hline
P.IDA.D3&	NIST SP 800-63A (Digital Identity Guidelines: Enrollment and Identity Proofing) \\ \hline
P.IDA.D4&	NIST SP 800-57 (Recommendation for Key Management)\\ \hline
P.PIA.D1&	\href{https://ico.org.uk/for-organisations/guide-to-data-protection/guide-to-the-general-data-protection-regulation-gdpr/accountability-and-governance/data-protection-impact-assessments/}{Data Protection Impact Assessment}\\ \hline
P.PIA.D3&	ISO/IEC 29134 (Guidelines for privacy impact assessment)\\ \hline
P.PIA.O1&	GDPR (General Data Protection Regulation) for transparency and communication requirements\\ \hline
P.PIA.O3&	ISO/IEC 27701 (Privacy Information Management System)\\ \hline
P.PIA.O4&	SO/IEC 27701 (Privacy Information Management System)\\ \hline
P.PIA.O6&	NIST Privacy Framework\\ \hline
P.PRMP.O5&	NIST SP 800-161 (Supply Chain Risk Management Practices for Federal Information Systems and Organizations)\\ \hline
P.WDPP.O4&	SO/IEC 29134 (Guidelines for privacy impact assessment)\\ \hline
P.SU.O2&	NIST SP 800-41 (Guidelines on Firewalls and Firewall Policy)\\ \hline
P.SU.O3&	NIST SP 800-192 (Verification and Test Methods for Access Control Policies/Models)\\ \hline
P.PS.D1&	NIST SP 800-63B (Digital Identity Guidelines: Authentication and Lifecycle Management)\\ \hline
P.PS.O1	&https://www.iso.org/standard/71670.html\\ \hline
P.PS.O2	&https://gdpr-info.eu\\ \hline
P.PS.O3	& \href{https://digital-strategy.ec.europa.eu/en/policies/cybersecurity-act}{EU Cybersecurity Act}\\ \hline
P.PS.O4	&https://www.iso.org/standard/72024.html\\ \hline
P.PS.O5	&https://www.iso.org/standard/45124.html\\ \hline
RES.RS.O1&	NIST SP 800-53(r4)\\ \hline
REL.SUS.D1&	\href{https://www.fatf-gafi.org/media/fatf/documents/recommendations/pdfs/Guidance-on-Digital-Identity-report.pdf}{Guidance on DiD}\\ \hline
REL.AL.D1	&ICAO 9303\\ \hline
REL.AL.O7	&ISO/IEC 29115:2013 \\ \hline
REL.AL.O8	&ISO/IEC TS 29003  \\ \hline
REL.AL.O9	&ISO 24760 51 \\ \hline
REL.AL.O10	&eIDAS\\ \hline
REL.GA.O1	&ISO/IEC 17065 (Conformity assessment -- Requirements for bodies certifying products, processes, and services)\\ \hline
REL.HUT.D1	&NIST SP 800-92 (Guide to Computer Security Log Management)\\ \hline
E.OP.D1	&ISO/IEC 18033 (Encryption algorithms)\\ \hline
E.OP.O1	& NIST SP 800-50 (Building an Information Technology Security Awareness and Training Program)\\ \hline 
\caption{Sample metrics and associated sources} 
\label{tab:standards}
\end{longtable}


 \bibliographystyle{elsarticle-num} 
 \bibliography{cas-refs}





\end{document}